\documentclass[prd,twocolumn,nofootinbib]{revtex4}

\usepackage{amsmath,amssymb,graphicx,subfigure}

\begin{document}

\preprint{\hepth{0712.3324}}

\title{Boltzmann babies in the proper time measure}

\author{Raphael Bousso, Ben Freivogel and I-Sheng Yang\footnote{bousso@lbl.gov, freivogel@berkeley.edu, 
jingking@berkeley.edu}}

\affiliation{Department of Physics and Center for Theoretical
Physics \\
University of California, Berkeley, CA 94720, U.S.A. \\
{\em and}\\
Lawrence Berkeley National Laboratory, Berkeley, CA 94720, U.S.A. }

\begin{abstract}
  After commenting briefly on the role of the typicality
  assumption in science, we advocate a phenomenological approach to
  the cosmological measure problem.  Like any other theory, a measure
  should be simple, general, well-defined, and consistent with
  observation.  This allows us to proceed by elimination.  As an
  example, we consider the proper time cutoff on a geodesic
  congruence.  It predicts that typical observers are quantum
  fluctuations in the early universe, or Boltzmann babies.  We sharpen
  this well-known youngness problem by taking into account the
  expansion and open spatial geometry of pocket universes.  Moreover,
  we relate the youngness problem directly to the probability
  distribution for observables, such as the temperature of the cosmic
  background radiation.  We consider a number of modifications of the
  proper time measure, but find none that would make it compatible
  with observation.
\end{abstract}

\maketitle

\section{Introduction}

\subsection{Typicality}

Every time we interpret an experiment, we assume that we are a typical
observer.  Suppose, for example, that we are trying to distinguish
between two theories $T_1$ and $T_2$.  Conveniently, they predict a
very different value of the spin of an electron subjected to a
suitable sequence of interactions: $T_1$ predicts spin up with
probability $\epsilon$, and $T_2$ predicts spin down with probability
$\epsilon$.  

If $\epsilon\ll 1$, then even a single measurement will allow us to
rule out one of these theories with considerable confidence.  We can
improve our confidence by repeating the experiment, but for
simplicity, let us suppose that $\epsilon$ is so miniscule that we are
satisfied with doing a single experiment.

In drawing the above conclusions, we acted as if our laboratory either
was the only laboratory in the universe, or was selected at random
from among all the laboratories doing the same experiment in the
universe.  This is the assumption of typicality.  Note that we have no
direct evidence for this assumption.  We do not know whether there are
other laboratories performing the same experiment on some far-away
planets; and if there are, then our laboratory was presumably not
actually selected by anyone from among them.  Nevertheless, the
overall success of the scientific method so far suggests that this
assumption is appropriate.

To see this, consider a prescription favored by Hartle and
Srednicki~\cite{HarSre07}, who decline to assume typicality.  They argue
that it does not matter whether a given outcome is likely to occur in
a randomly chosen laboratory; what matters is whether one is likely to
be able to find {\em some\/} laboratory, somewhere in all of
spacetime, no matter how atypical, in which that outcome occurs.  This
probability is given not by $\epsilon$, but by $1-(1-\epsilon)^L$,
where $L$ is the number of laboratories in the universe.

The effect of using this probability-of-global-existence is most
dramatic in the case where $L\gg \epsilon^{-1}\gg 1$.  Then we cannot
rule out either theory, no matter what we observe.  We can still rule
out one of the two theories by repeating the experiment sufficiently
often.  But to know at which point we can reject one of the theories,
we would need to know how many other laboratories there are.  Since we
do not know $L$, the Hartle-Srednicki prescription would put an end to
experimental science.  It would render all experiments pointless,
because we could not reject any theory until we know how many other
laboratories there are.  Given the success of the scientific method
thus far, we may conclude the Hartle-Srednicki prescription is
inappropriate.\footnote{We cannot conclude that our laboratory is the
  only one, since we could simply build a second one.  Note that in
  the Hartle-Srednicki prescription, this would be inadvisable, since
  it would render the experiments performed at either laboratory less
  conclusive.}

Here we have argued for the assumption of typicality on empirical
grounds: it has served us well as a heuristic tool.  If it was wrong,
we should not have been successful in devising and rejecting
scientific theories on the basis of this assumption.  But {\em why}\/
does it work so well?  This, too, can be understood; elegant
discussions have recently been given by Page~\cite{Pag07}, and by
Garriga and Vilenkin~\cite{GarVil07}, who also offer a careful
definition of the class of observers among which we may consider
ourselves to be typical.

\subsection{The measure problem: a phenomenological approach}

In the multiverse, we can use typicality to make statistical
predictions for the results of observations. For instance, to predict
the cosmological constant, we would first determine the theoretically
allowed values, and then count the number of observations of each
value.  The probability to observe a given value of the cosmological
constant is proportional to the number of observations, in the
multiverse, of that value.  The problem is that under rather generic
conditions, the universe will have infinite spacetime volume, even if
it is spatially finite (i.e., contains a compact Cauchy surface).
Then the number of observations can diverge.

The landscape of string theory contains perhaps $10^{500}$ metastable
vacua, allowing it to solve the cosmological constant
problem~\cite{BP}; see Refs.~\cite{Pol06,TASI07} for a review.
However, divergences would arise even if there was only one false
vacuum.  For example, suppose that there was a first-order phase
transition in our past, by which a long-lived metastable vacuum
decayed.  The symmetries of the instanton mediating this
decay~\cite{CDL} dictate that the resulting true vacuum region is an
infinite open FRW universe.  It will contain either no observers, or
an infinite number of them.  Moreover, the parent vacuum will keep
expanding faster than it decays, so that an infinite number of true
vacuum bubbles (or ``pocket universes'') are created over
time~\cite{GutWei83}.

The measure problem in cosmology is the question of how to regulate
these infinities, in order to get a finite count of the number of
observations of each type.\footnote{In general, the question of what
  constitutes one observation is a difficult problem.  For instance,
  it is not obvious precisely how many observations of the CMB
  temperature should be assigned to our local efforts on Earth.  (See
  Ref.~\cite{Bou06} for a recent proposal using entropy production.)
  However, these considerations are orthogonal to the issue at hand,
  which is the regularization of the infinite spacetime four-volume
  arising in eternal inflation.  Here, we will assume that the local
  counting of observations is unambiguous.}  The choice of measure is
no minor technicality, but an integral part of a complete theory of
cosmology.  Two different measures often assign exponentially
different relative probabilities to two types of
observations.\footnote{A simple example arises even if there is only
  one false vacuum.  Each true vacuum bubble collides with an infinite
  number of other such bubbles, so one may ask whether we are likely
  to live in a collision region.  Leaving aside fatal effects of
  collisions, this probability is nevertheless exponentially small in
  the proper time measure considered here, and also in the causal
  diamond measure~\cite{Bou06}.  But if one averages over worldlines
  emanating from the nucleation point, all but a set of measure zero
  of them will immediately enter a collision region~\cite{GarGut06}.}


Ultimately, a unique measure should arise from first principles in a
fundamental theory~\cite{FreSus04,FreSek06,Sus07,MalSheSus}.  
In the meantime, however, we may regard the measure problem as a {\em
phenomenological\/} challenge.  At least in the semiclassical
regime, we can hope to identify the correct measure by the traditional
scientific method: We try a simple, minimal theory, and work out its
implications.  If they conflict with observation, we either refine
(i.e., complicate) the model, or we abandon it altogether for a
different approach.

What one may regard as a simple measure is, to some extent, in the eye
of the beholder.  The same can be said for simple theories; yet, for
the most part, we know one when we see one.  Only a handful of
measures have been proposed (see, e.g, Refs.~\cite{Vil06,Lin06,Van06}
for overviews and further references),
and many of them can be seen to conflict with observation, often
violently.  This is good news, because it makes it feasible to proceed
by elimination.  Let us investigate simple proposals, let us ask
whether they are well-defined, and let us determine whether they
conflict with observation.

For example, consider the proposal of Ref.~\cite{GarSch05}.  In its
original form, it predicted with probability 1 that we should find
ourselves as isolated observers (``Boltzmann brains'') resulting from
a highly suppressed thermal fluctuation in a late, empty
universe~\cite{Pag06,BouFre06b}.  This led to a
refinement~\cite{Vil06b}, which complicates the measure and seems {\em
  ad hoc}~\cite{Pag06b}.  Depending on the details of the string
landscape, the proposal may render most vacua dynamically inaccessible
(the ``staggering problem'' of Refs.~\cite{SchVil06,OluSch07}).  This
would also amount to a conflict with observation, namely the
prediction that we should observe a much larger cosmological constant
with probability very close to 1.  Perhaps most importantly, at
present the proposal is well-defined only in the thin-wall limit of
bubble formation, and if bubble collisions are
neglected~\cite{GarGut06}.\footnote{The proposal cuts off the infinite
  number of observers in different vacuum bubbles by restricting to a
  ``unit comoving volume'', defined by appealing to the universality
  of the open universe metric inside every bubble at early
  times~\cite{GarSch05}.  But universality holds only if the thickness
  of the wall, and its collisions with other bubbles, are both
  neglected.  These two assumptions cannot both be satisfied to a good
  approximation.}

Another recent proposal~\cite{Bou06}, the ``holographic'' or ``causal
diamond'' measure, has so far fared well.  It is
well-defined in the semiclassical limit, and it does not have a
staggering problem~\cite{BouYan07}.  Its prediction of the
cosmological constant agrees significantly better with the data than
that of any other proposal~\cite{BouHar07}, and it continues to agree
well even as other parameters are allowed to vary~\cite{CliFre07}.  It
will be important to test this proposal further, for example, by
allowing even more parameters to vary.  But it is encouraging that we
have at least one well-defined measure that has not been ruled out.

In this paper, we consider a much older proposal, the {\em proper time
  measure\/}~\cite{Lin86a,LinLin94,GarLin94,GarLin94a,GarLin95}.  At
present, this measure is not completely well-defined, and we will
comment on some issues that will have to be overcome to make it
well-defined.  But our main focus will be on its well-known conflict
with observation, the ``youngness paradox''.  In particular, we will
investigate whether simple modifications of the measure can resolve
this problem.

\subsection{The proper time measure and the youngness paradox}

To apply the proper time measure, one begins by selecting an (almost
arbitrary) finite portion of a spacelike slice in the semiclassical
geometry.  The congruence of geodesics orthogonal to this initial
surface defines Gaussian normal coordinates, and thus a time slicing,
at least until caustics are encountered. The number of observations
between the initial slice and the time $t$ is finite.  Globally, the
multiverse reaches a self-reproducing state at late times: its volume
expands exponentially, but the ratio of different types of
observations remains constant and finite.  Therefore, relative
probabilities defined by this measure are independent of the initial
conditions.

Earlier work~\cite{LinLin96,Gut00a,Gut00b,Gut04,Teg05,Lin07,Gut07} has 
already shown that the proper time measure has a youngness problem: it
predicts with essentially 100\% probability that we should be living
at an earlier time.  The reason for this problem can roughly be
described as follows.

The asymptotic rate of expansion of the multiverse is dominated by the
vacuum with the largest Hubble constant $H_{\rm big}$, which defines a
microphysical time-scale $H_{\rm big}^{-1}$.  (In the string
landscape, this would be of order the Planck time.)  For simplicity,
let us consider only regions occupied by our own vacuum.  We may ask
about the distribution of the age of such bubbles, i.e., how long
before the cutoff $t$ they were formed.  In particular, we may ask how
many bubbles are at least $13.7$ Gyr old, and thus contain
observations like ours; and we may compare this to the number of
bubbles that are, say, $13$ Gyr old.  The size of the bubble interior
is not much affected by these different time choices, but the number
of bubbles will be vastly different.  For every bubble that is at
least $13.7$ Gyr old at the time $t$, there will be of order $\exp
\left(3\times 0.7 \,{\rm Gyr}/ H_{\rm big} ^{-1}\right)$ bubbles that
are $13$ Gyr old, because of the overall exponential growth of the
volume of the multiverse in the extra 700 million years before it has
its last chance to nucleate the younger bubbles.  Perhaps the younger
bubbles contain fewer observers per bubble, but surely not so few as
to compensate for a factor $\exp(10^{60})$.  This mismatch persists as
$t\to\infty$. Thus, typical observers are younger than we are, and the
probability for an observer to live as late as we do is
$\exp(-10^{60})$.  This rules out the proper time measure at an
extremely high level of confidence.

Of course, our choice of $13$ Gyr observers as a comparison group is
arbitrary.  Because $H_{\rm big}^{-1}$ is a microphysical scale, even
observers just one minute younger (relative to their big bang) are
superexponentially more probable than we are.  Ultimately, one should
consider observers of any cosmological age.  Because of the
exponential pressure to be young, it pays to arise from a rare quantum
fluctuation in the early universe.  The most likely observers are such
``Boltzmann babies'', and the most likely observations are the
phenomena of the hot, dense, early universe they see.

\subsection{Summary and outline}

Our goal in this paper is two-fold.  First, we will make the youngness
paradox more precise.  Traditional treatments have neglected the
expansion of new bubbles.  We supply a justification for this ``square
bubble'' approximation, by extending Gaussian normal coordinates
across an expanding bubble wall, and showing that our exact treatment
reproduces the usual youngness problem.  We will distinguish
carefully between probability distributions for the {\em time\/} when
observers live (which is not directly observable), and probability
distributions for actual observables, like the temperature of the
background radiation measured by observers~\cite{Teg05}.  We find that
the youngness paradox manifests itself by predicting that we should
observe a higher temperature than 2.7 K, with probability
exponentially close to 1.

Our second goal is to consider possible modifications of the proper
time measure.  We will argue that it is difficult to resolve the
youngness paradox, other than by abandoning the measure altogether.
In particular, Linde has proposed a modification in the context of a
particular toy model~\cite{Lin07}. Since no general prescription was
given, it is not clear how to extend this modification to other
settings, and in particular to the probability distribution for the
observed background temperature.  We consider a number of possible
choices, some of which reduce to the prescription of Ref.~\cite{Lin07}
for the particular probabilities computed therein.  However, we are
unable to find any modification that escapes all of the conflicts with
observation that arise from the youngness problem.  In particular, 2.7
K remains an extremely atypical value of the background temperature
under all choices we consider.

The structure of the paper is as follows.  In Sec.~\ref{sec-proper},
we explain the proper time measure in more detail. In
Sec.~\ref{sec-geod}, we compute the paths of geodesics entering
bubbles, in order to determine the shape of the proper time cutoff
within bubbles. In Sec.~\ref{sec-prob}, we compute the probability
distribution for the spacetime location of observers, finding a
youngness paradox and conflict with observation.  In
Sec.~\ref{sec-fixes}, we try a few modifications of the measure, but
find no simple modification consistent with observation.

\section{The Proper Time Measure}
\label{sec-proper}

The proper time measure (sometimes referred to as the ``standard
volume weighted measure'') is one of the simplest and most
straightforward ways of regulating the infinities of the multiverse.
Choose a small three-dimensional patch of space, $\Sigma_0$,
orthogonal to at least one eternally inflating geodesic.  Then,
construct Gaussian normal coordinates~\cite{Wald} in its future.  That
is, a given event has the time coordinate $t$ if it occurs at proper
time $t$ along a geodesic orthogonal to $\Sigma_0$.  Such events form
a three-dimensional hypersurface $\Sigma_t$.  The regularization
scheme is to count only observations between proper time hypersurfaces
$\Sigma_0$ and $\Sigma_t$.  Relative probabilities are defined by
ratios, in the limit $t\rightarrow\infty$.

\begin{figure}[h!]
\begin{center}
\includegraphics[viewport=600 340 0 0,width=8 cm,clip,angle=180]{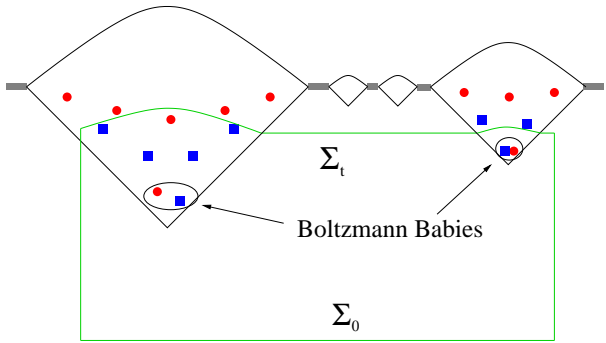}
\caption{The relative probability of making different observations,
  for example two different CMB temperatures (red disks or blue
  boxes), is determined by simple counting in the finite region
  between $\Sigma_0$ and $\Sigma_t$.  The ratio tends to a finite
  limit as $t\rightarrow\infty$.  The youngness problem is the fact
  that anomalous early fluctuations producing either observation
  (Boltzmann babies) turn out to dominate the count.  To show this
  correctly in the figure, one would need to draw an exponentially
  large number of ``young bubbles'', like the one on the right, in
  which only the Boltzmann babies contribute.}
\label{fig-multi}
\end{center}
\end{figure}

It is well-known that Gaussian normal coordinates are only locally
defined.  They break down at {\em caustics}, or focal points, where
infinitesimally neighboring geodesics in the congruence intersect.
Beyond such points, the above definition of the time coordinate $t$ is
ambiguous.  We sidestep the issue here by considering only expanding
spacetime regions and ignoring clustering and inhomogeneities (and
thus, strictly speaking, all known observers), so that focusing does
not occur.

Let $O_1$ and $O_2$ be two mutually exclusive observations.  For
example, $O_1$ may subsume any observation made in vacuum $A$, while
$O_2$ corresponds to vacuum $B$.  Or $O_1$ ($O_2$) may be capture
information about the observer's spatial or temporal location within a
given vacuum, for example the fact that universe is matter (vacuum)
dominated.

Let $N_i(t)$ be the number of observations of type $O_i$ made in the
four volume between $\Sigma_0$ and $\Sigma_t$.  Observations
take a finite time, so for definiteness let us demand that an
observation must be complete for it to be counted.  The relative
probability for the two observations is defined to be
\begin{equation}
\frac{p(O_1)}{p(O_2)}=\lim_{t\to\infty}\frac{N_1(t)}{N_2(t)}~.
\label{eq-prob}
\end{equation}

Similarly, we can consider a continuous set of possible observations
$O_T$, such as the observation of a CMB temperature $T$.  In this
case, we are interested in the probability density $dp/dT$, which is
given by
\begin{equation}
  \frac{\left.\frac{dp}{dT}\right|_{T_1}
  }{\left.\frac{dp}{dT}\right|_{T_2}}=
  \lim_{t\to\infty}
  \frac{\left.\frac{dN}{dT}\right|_{T_1}(t)
  }{\left.\frac{dN}{dT}\right|_{T_2}(t)}
\label{eq-density}
\end{equation}
Here, $\left.\frac{dp}{dT}\right|_{T_1}dT$ is the
probability of observing $T$ in the interval
$(T_1,T_1+dT)$.
$\left.\frac{dN}{dT}\right|_{T_1}dT$ is the number of
instances of such observations in the four-volume between $\Sigma_0$
and $\Sigma_t$.\footnote{In the continuous case, one must take care
  with the order of limits.  First we pick a finite $dT$ and define
  ratios as usual, by taking $t\to\infty$.  Then we repeat the
  procedure while taking $dT\to 0$.}

At late times, $N_i(t)\propto \exp(3H_{\rm big}t)$.  The overall
scaling rate $H_{\rm big}$ is set by the most rapidly expanding
vacua~\cite{Linde}.  [In the string landscape, one expects that
$H_{\rm big}\sim {O}(1)$ in Planck units.]  This exponential
growth guarantees that
\begin{equation}
  \lim_{t\to\infty}\frac{N_1(t)}{N_2(t)}=
  \lim_{t\to\infty}\frac{N_1'(t)}{N_2'(t)}~,
\end{equation}
where $N_i'=dN_i/dt$ is the rate at which observations of type $O_i$
are being made, integrated over space but not over time.  Thus, it
does not matter whether probabilities are computed from the total
number of observations until the time $t$, or the rate of observations
at the time $t$, or the number of observations made in some recent
(fixed width) time interval $(t-\Delta t,t)$.  For definiteness,
however, we will stick to the first of these definitions.

\section{Geodesics crossing bubbles}
\label{sec-geod}
\subsection{Open FRW time vs. geodesic time}

\begin{figure}[t!]
\begin{center}
\includegraphics[scale=1]{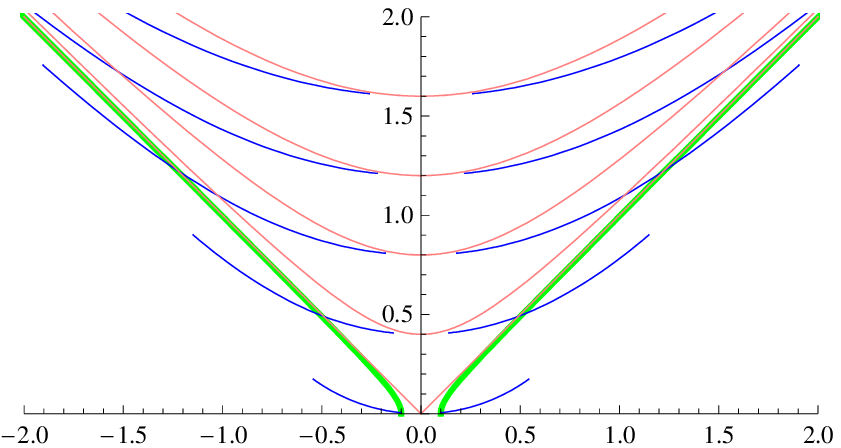}
\includegraphics[scale=1]{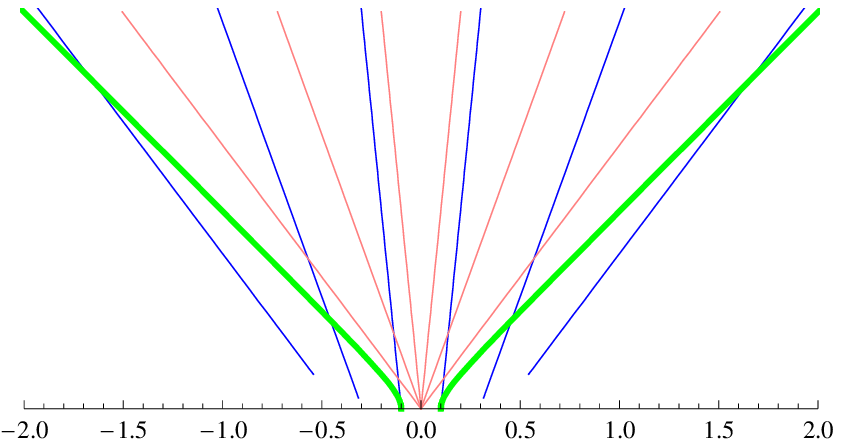}
\caption{The top figure shows slices of constant FRW time, $\tau$
  (red, light) and slices of constant geodesic time $t$ (blue, dark)
  in the vicinity of a bubble wall (green, thick) with initial size
  $r_0=0.1\, H_{\rm out}^{-1}$.  Note that the constant $t$ slices are
  not defined for geodesics passing through the nucleation region of
  the bubble.  The lower figure shows that geodesics of the congruence
  (blue, dark) eventually asymptote to comoving FRW worldlines (red,
  light).}
\label{fig-slic}
\end{center}
\end{figure}

The measure discussed above was first applied to slow-roll models of
eternal inflation, without first-order phase
transitions~\cite{LinLin94,GarLin94}.  In this case, one keeps track
of fluctuations of scalar fields on the Hubble scale, effectively
assuming that they decohere every Hubble time (see
Ref.~\cite{BouFre06a} for a discussion of the validity of this
approach).  There is no obstruction to applying the same measure to
models with bubble formation, but there is an annoying complication
(Fig.~\ref{fig-slic}).

Consider a region of the universe at late times, occupied by a
``host'' de~Sitter vacuum with cosmological constant $3H_{\rm out}^2$.
Let us suppose that $H_{\rm out}$ is very large, but far enough below
the Planck scale to lend validity to our semiclassical treatment.
Moreover, we suppose that a bubble of our own vacuum can form by a
Coleman-DeLuccia (CDL) tunneling process inside the host vacuum.

Let us suppose, moreover, that the host vacuum has existed for many
Hubble times $H_{\rm out}^{-1}$.  Then, on the scale about to be
occupied by a newly formed bubble, the geodesics emanating from the
initial surface $\Sigma_0$ can be treated as comoving in the flat
de~Sitter metric
\begin{equation}
ds^2=-dt^2+ H_{\rm out}^{-2} e^{2H_{\rm out}t} d {\mathbf x} ^2
\label{eq-flat}
\end{equation}
This follows, in a sense, from the de~Sitter no-hair theorem; we will
also find that it is consistent with our careful analysis in
Sec.~\ref{sec-exact}. 

Now suppose that a bubble of our vacuum forms at the time $t_{\rm nuc}
$.  It will appear at rest, with a proper radius $r_0\ll H_{\rm
  out}^{-1}$ determined by the CDL instanton.  Then it will expand at
constant acceleration $r_0^{-1}$, its world-volume asymptoting to a
light-cone.  Some of the above geodesics will eventually run into the
bubble wall and enter our universe.  Their behavior will
determine the weight of any observations carried out inside the
bubble, in the proper time measure.

We will not consider the
  small subset of geodesics that go through the nucleation region,
  $r\lesssim r_0$, where the classical geometry is not clearly
  defined.  It is unclear how to treat these geodesics.  This
  constitutes a challenge for the sharp formulation of
  congruence-based measures.  The best we can say is that our results
  show that values of $r\sim O(r_0)$ contribute negligibly to the
  measure as they are approached from above in a controlled regime.
  This could be viewed as evidence that the contribution of the
  uncontrolled regime can also be neglected.

The metric inside the bubble is given by an open FRW geometry;
ignoring fluctuations, the metric is
\begin{equation}
ds^2=-d\tau^2+a(\tau)^2 (d\xi^2+\sinh^2\xi\, d\Omega_2^2)~.
\label{eq-FRW}
\end{equation}
where the scale factor $a(\tau)$ comprises, for anthropically
relevant bubbles, a period of
inflation followed by radiation, matter, and vacuum domination with
very small cosmological constant.  Note that
$a(\tau)\approx \tau$ for sufficiently small $\tau$.

The maximally symmetric and negatively curved spatial slices
defined by $\tau = const$ are physically preferred inside the bubble, since they
correspond to hypersurfaces of (approximately) constant density.  Only
at very late times, well into the vacuum-dominated era, do we lose
this preferred slicing, as the universe again becomes locally empty
de~Sitter.

The key point is that the preferred surfaces of constant FRW time $\tau$
 are {\em  not\/} the surfaces  $\Sigma_t$ of constant geodesic time $t$.
    This is a complication, since
$\tau$ is what we usually call the age of the universe, the time since
the big bang---really, the metric distance from the bubble nucleation
event.  To the extent that any time variable is directly correlated
with the outcome of an observation (such as CMB temperature or the
amount of clustering), that variable will be the FRW time $\tau$, 
and not the global time $t$.

\subsection{Square bubble approximation}
\label{sec-square}

The {\em square bubble approximation\/}, which is implicit in
Ref.~\cite{Lin06}, aims to circumvent this complication.  It
amounts to a deformation of the metric that allows us to calculate as
if constant $\tau$ slices, inside the bubble, coincide with constant
$t$  slices.  For this we must arrange that the FRW
time $\tau$ and the geodesic time $t$ differ only through a
constant shift,
\begin{equation}
\tau=t-t_{\rm nuc}~.
\end{equation}
This is possible only if the movement of the bubble wall is neglected.

Given this {\em ad-hoc\/} modification, the continuation of the
geodesic congruence into the bubble cannot be directly computed.  We
will simply assume that the internal geometry of the new vacuum is a
spatially finite piece of a {\em flat\/} FRW universe
\begin{equation}
  ds^2=-d\tau^2 + \tilde a(\tau)^2 d {\mathbf y} ^2~.
\label{eq-insideflat}
\end{equation}
To match at $t=t_{\rm nuc} $ ($\tau=0$), we let $ {\mathbf y}$ range
over a finite physical volume $\tilde a(0)^3 V_y$.  We take the
comoving volume to be independent of $\tau$, as if the bubble wall
remained at fixed ${\mathbf y} $.

Note that both the scale factor and the initial size of the bubble
initially differ significantly from their true values, and the
matching to the outside fails at late times.  However, in inflating
vacua the exponential internal growth is more important than the
expansion of the bubble forming their boundary.  Moreover, inflation
locally washes out the difference between a flat and an open universe.
After a short time (say, a few e-foldings of inflation) we can take
$\tilde a(\tau)\propto a(\tau)$.

Nevertheless, the square bubble approximation blatantly contradicts
important known features, such as the fact that the constant-density
slices inside are actually open and infinite.  Indeed it is not even
consistent geometrically, making it impossible to match the inside of
the bubble to the outside.  But one may hope that it gives a
reasonable approximation {\em for the purpose of computing
  probabilities}.  This will be the case if the approximation does not
change the true count of observations of various types. We will find
that the square bubble approximation is a good one for many questions.

\subsection{Exact relation}
\label{sec-exact}
The actual relation between the FRW coordinates $(\tau,\xi)$ and the
geodesic proper time $t$ is more complicated. We set
\begin{equation}
H_{\rm out} = 1 \ \ {\rm in\ this\ subsection,}
\end{equation}
 so that the equations are not quite so ugly.
In the outside flat deSitter slicing, 
\begin{equation}
ds^2 = - dt^2 + e^{2 t } \left(dr^2 + r^2 d \Omega_2^2 \right) ~,
\end{equation}
 the domain wall 
follows the trajectory
\begin{eqnarray}
r_w \exp (t_w - t_{\rm nuc})&=&r_0\cosh\eta~, \\
\exp (t_w - t_{\rm nuc})&=&
r_0\sinh\eta+\sqrt{1-r_0^2}~. \label{eq-tw}
\end{eqnarray}  
Here $r_0$ is the size of the bubble at nucleation; it is also the
radius of curvature and the inverse proper acceleration of the domain
wall.  $r_0 \eta$ is the proper time along the domain wall.

We need to compute the motion of geodesics as they cross the domain
wall and live happily ever after in the interior. The natural
coordinates inside the bubble are the open FRW coordinates $(\tau,
\xi)$ of Eq.~(\ref{eq-FRW}) because they respect the symmetry of the
bubble nucleation. However, these coordinates do not cover the region
containing the domain wall, so it is convenient to use a different
coordinate system near the domain wall.  Assuming that the Hubble constant 
in the interior of the bubble is much smaller than the Hubble constant in 
the exterior, we can find a scale $\tau^*$ such that 
$H_{\rm in}^{-1}\gg\tau^*\gg H_{\rm out}^{-1}$.  
The region from the domain wall to the 
$\tau^*$ surface is much smaller than the characteristic scales of the geometry inside the bubble.  As a result, we can approximate it as a piece of Minkowski space. We will use coordinates in which the metric is
\begin{equation}
ds^2 = - dT^2 + dR^2 + R^2 d\Omega_2^2~.
\end{equation}
 
Because the domain wall is a constant curvature surface with curvature
radius $r_0$, its trajectory in the Minkowski coordinates is
\begin{eqnarray}
R_w&=&r_0\cosh\eta~, \\
T_w&=&r_0\sinh\eta~,
\end{eqnarray} 
where again $r_0 \eta$ is the proper time along the domain wall.
Computing the 4-velocity, we find that $\eta$ is the rapidity of the
domain wall.
  
The trajectory of the geodesic after crossing the domain wall is
\begin{eqnarray}
T&=&r_0\sinh\eta+(t-t_w)\cosh\alpha \label{eq-T}~, \\
R&=&r_0\cosh\eta+(t-t_w)\sinh\alpha \label{eq-R}~,
\end{eqnarray}
where $\alpha$ is the rapidity of the geodesic.  We will determine
$\alpha$ by demanding that the angle between the domain wall and the
geodesic is continuous across the domain wall\footnote{At the domain
  wall, the first derivative of the metric is discontinuous, resulting
  in a delta function in $R_{\mu\nu}$ in the thin wall
  approximation. However because the connection only depends on first
  derivatives, we can find a local coordinate system where the
  connection $\Gamma^\gamma_{\mu\nu}$ is finite.  This proves that the
  angle (inner product) between the geodesic and the domain wall is
  continuous across the wall.}. If $u$ and $v$ are the 4-velocities of
the geodesic and the domain wall, we demand
\begin{equation}
u\cdot v|_{\rm out} =u\cdot v|_{\rm in} ~.
\end{equation}
Since $\alpha$ is the rapidity of the geodesic and $\eta$ is the
rapidity of the domain wall,
\begin{equation}
u\cdot v|_{\rm in} =\cosh(\eta - \alpha)~.
\end{equation}
Geodesics outside the domain wall have a simple 4-velocity $u_{\rm
  out} = (1, 0, 0, 0)$, and since we have identified $r_0 \eta$ as the
proper time along the domain wall, the 4-velocity of the domain wall
is $v = ({1 \over r_0} ~ d t_w /d \eta, ...)$. Using the equation
(\ref{eq-tw}) for the trajectory of the domain wall,
\begin{equation}
u\cdot v|_{\rm out}
={1 \over r_0} \frac{d t_w }{d \eta}
=\frac{\cosh\eta}{r_0\sinh\eta+\sqrt{1-r_0^2}}~.
\end{equation}
Thus the equation determining $\alpha$ is
\begin{equation}
\frac{\cosh\eta}{r_0\sinh\eta+\sqrt{1-r_0^2}}
=\cosh(\eta-\alpha)~.
\label{eq-angle}
\end{equation}

It is convenient to combine Eq.~(\ref{eq-angle}) and (\ref{eq-tw})
to get
\begin{equation}
\cosh \eta \exp \left[-(t_w - t_{\rm nuc})\right] = \cosh(\eta - \alpha)~.
\end{equation}
Simplifying we find
\begin{equation}
(t_w - t_{\rm nuc})=\alpha-\ln(1+\varepsilon)~,
\label{eq-raptw}
\end{equation}
where
\begin{equation}
\varepsilon=\frac{e^{2 \alpha} - 1}{e^{2 \eta} + 1}~.
\label{eq-eps}
\end{equation}
This is a convenient rewriting because one can show that $\varepsilon
\ll 1$ for all geodesics as long as the critical bubble size is small
in Hubble units, $r_0 \ll 1$.

We want to rewrite the geodesics in terms of the open FRW coordinates
which will be adapted to the cosmological evolution inside the
bubble. For $\tau\ll H_{\rm in}^{-1}$, where the geometry is approximately
Minkowski space, the relationship is
\begin{eqnarray}
\tau &=& \sqrt{T^2-R^2} \label{eq-tau} \\
\xi  &=& \tanh^{-1}\frac{R}{T} \label{eq-xi}~.
\end{eqnarray}
Using the trajectory in $(R,T)$ given by Eq.~(\ref{eq-T}), (\ref{eq-R}), 
we find the trajectory in FRW coordinates
\begin{eqnarray}
\xi&=&\alpha+{O}(\frac{1}{\tau})~,
\label{eq-xia}
\\
\tau &=&\sqrt{(t - t_w)^2+2r_0(t-t_w)\sinh(\eta-\xi)-r_0^2}~.
\label{eq-taua}
\end{eqnarray}
Our goal is to manipulate all of the above equations in order to find
a single equation for the geodesic time since nucleation, $ t - t_{\rm
  nuc}$, as a function of the natural coordinates $\tau, \xi$ inside
the bubble.

We will be interested in events which occur a reasonable distance away
from the domain wall, so that $\tau, t-t_{\rm nuc}, t-t_w \gg 1 $. So
we can drop the subleading terms in (\ref{eq-xia}) and just set the
rapidity of the geodesic equal to the comoving coordinate, $\alpha =
\xi$. Physically, the point is that the final comoving position of the
geodesic is determined only by its velocity and not by its initial
location. The nontrivial statement in (\ref{eq-xia}) is that the
geodesics become comoving in a time set by the Hubble scale outside
the bubble, $H_{\rm out}^{-1}$.
 
In Eq.~(\ref{eq-taua}) we can now set $\alpha = \xi$ and expand for
large $t-t_w $ to get
\begin{equation}
\tau =t-t_w + r_0 \sinh(\eta - \xi)~.
\end{equation}
Going back to (\ref{eq-angle}) and solving for $\sinh(\eta - \xi)$ we find
\begin{equation}
\sinh(\eta - \xi) = \frac{\sqrt{1-r_0^2}\sinh\eta-r_0}
{ r_0 \sinh\eta+\sqrt{1-r_0^2}}~.
\end{equation}
Using the relation (\ref{eq-tw}) between $\eta$ and $t_w$ this can be
rewritten as
\begin{equation}
  \sinh(\eta - \xi) = {1 \over  r_0} \left[\sqrt{1 -  r_0^2}
    - \exp \left[-(t_w-t_{\rm nuc})\right] \right]~.
\end{equation} 
So we have an equation relating the geodesic time since nucleation to
the FRW time $\tau$ and the time $t_w$ the geodesic crosses the domain
wall:
\begin{equation}
  \tau = t - t_w +   \left[\sqrt{1 -  r_0^2} - 
    \exp[-(t_w-t_{\rm nuc})] \right]~.
\end{equation}

Now we can use the relation (\ref{eq-raptw}) between the rapidity
$\alpha$ and $t_w$, together with $\alpha= \xi$, to get
\begin{equation}
\tau = t - t_{\rm nuc}-\bigg[\xi + e^{- \xi} -
\sqrt{1-r_0^2}+ \varepsilon e^{-\xi} -\ln(1+\varepsilon)\bigg]~,
\label{eq-exact}
\end{equation}
where $\varepsilon$ is given by (\ref{eq-eps}).  Expanding in
$\varepsilon$ and restoring the factors of $H_{\rm out}$, we get the
final formula relating the geodesic time to the natural coordinates
inside the bubble:
\begin{equation}
  t - t_{\rm nuc} = \tau + H_{\rm out}^{-1}
  \left[ \xi + e^{- \xi} - 1 + ... \right]~.
\label{eq-app}
\end{equation}
As expected, the difference between the geodesic proper time and the
open FRW time depends non-trivially on the radial FRW coordinate
$\xi$.

\section{The spacetime location of a typical observer}
\label{sec-prob}

The proper time measure makes nontrivial and interesting predictions
for vacuum selection, which do not appear to contradict anything we
know~\cite{CliShe07}.  However, as soon as we ask about the
probabilities of different observations in the same vacuum, the
measure wildly conflicts with observation.  It has two properties that
result in a squeeze.  On the one hand, for an observation to be
counted, it must occur before the cutoff $t$.  On the other hand, the
multiverse as a whole is expanding exponentially on a microscopic
characteristic time scale.  This makes it favorable to wait as long as
possible until creating a low-energy, slowly expanding region like the
one in which we are making our observations, and it strongly favors
observations that happen soon after the fastest expanding vacuum has
decayed.  This is the general idea of the youngness paradox
~\cite{LinLin96,Gut00a,Gut00b,Gut04,Teg05,Lin07,Gut07}.

We will present one explicit calculation to show the fact that, within 
bubbles identical to ours, the probability to live at $13.7$ Gyr is 
vanishingly small compared to the probability to live at $13$ Gyr.  
There is nothing new in this calculation, but it will be easier to see
how the exact geometry we found goes into the paradox in Sec.
\ref{sec-bubble}, and how to analyze possible modifications in Sec.
\ref{sec-fixes}.  

Another manifestation of the youngness paradox is that if a number of
tunneling events are necessary to get from the fastest inflating
vacuum to our host vacuum, these successive tunneling events will tend
to be separated by only the Planckian time interval $H_{\rm
  big}^{-1}$. Since the tunneling events are not well-separated, this
renders it difficult to compute semiclassically. However, since such a
quick succession of tunneling events does not obviously contradict
observation, we sidestep this difficulty here by assuming that our
vacuum is produced directly from the fastest inflating vacuum. Hence
we set 
\begin{equation}
H_{\rm out} = H_{\rm big}~.
\end{equation}
This simplification makes the problem more well-defined; however,
every indication is that the characteristic time scale appearing in
the youngness paradox is $H_{\rm big}^{-1}$, regardless of this
simplification.

\subsection{The youngness paradox in the square bubble approximation}

We begin with an analysis in the square bubble approximation defined
in Sec.~\ref{sec-square}.  By assumption, each bubble appears as a
flat patch of the same physical size.  Hence, the comoving volume
$V_x$ taken up by a bubble in the outside metric goes like
$\exp(-3H_{\rm big} t_{\rm nuc})$.  Note, however, that we have rescaled
the Euclidean spatial coordinates inside the bubble, $d{\mathbf y}=\exp(
  H_{\rm big} t_{\rm nuc}) d{\mathbf x}$, so as to make the metric in
each bubble explicitly the same (not just equivalent by
diffeomorphism).  In the sequel, ``comoving volume'' will usually
refer to the inside metric and will accordingly be denoted $V_y$.  It
will be the same for each bubble, and so will drop out of
probabilities.

Let $n(t)$ be the total number of bubbles of our type produced prior
to the time $t$.  This grows exponentially with time:
\begin{equation}
  n(t)=C \exp(3 H_{\rm big}t)~.
\label{eq-nour}
\end{equation}
Here $C$ is a fixed constant, which depends on the size of $\Sigma_0$,
the initial state, and the rate at which our vacuum is produced
(directly or indirectly) by the fastest inflating vacuum.  This
constant will drop out in all ratios.

The nucleation of a bubble like ours will be followed by the formation
of observers.  Let $dN^{(1)}$ be the number of observations of some
type, in a single bubble of our type, in a comoving volume of size
$dV_y$ during the proper time interval $(\tau, \tau+d\tau)$ after the
formation of the bubble.  By the homogeneity of the FRW universe,
$dN^{(1)}$ will depend only on the FRW time, $\tau$, so we can write
\begin{equation}
  dN^{(1)} = f(\tau)\, d\tau \, d V_y ~.
\end{equation}
The function $f(\tau)$ can be thought of as an observer density.

As long as these observations involve looking out into the sky, they
will usually be different at different times $\tau$.  For simplicity,
we begin by treating $\tau$ itself as an ``observable'', and computing
the probability density
\begin{equation}
  \frac{dp}{d\tau}\propto
  \lim_{t\to\infty}\frac{dN}{d\tau}(t)~.
\end{equation}
This is the probability for an observer to find themselves living a
time $\tau$ after the big bang of their bubble.

Both the observer distribution, $f(\tau)$, and the volume per
bubble, $V_y$, are the same for all bubbles of our type, by the above
assumptions.  Therefore, the total number of $\tau$-observations
made by the time $t$ depends on $t$ only through the total number of
bubbles $n(t-\tau)$ produced prior to the time $t-\tau$:
\begin{eqnarray}
  \frac{dN}{d\tau}(t) &=& 
  n(t - \tau)   \frac{dN^{(1)}}{d\tau} (\tau)
  =
  f(\tau)\, V_y\, n(t-\tau) \nonumber \\ 
  &=& f(\tau)\, V_y\, \exp \left[3 H_{\rm big} (t-\tau) \right]~.
\end{eqnarray}
Since the $t$ dependence of the answer is just an overall
normalization, it drops out of the probability distribution and we get
the simple answer
\begin{equation}
 \frac{dp}{d\tau}\propto f(\tau) \exp(-3H_{\rm big}\tau) ~.
 \label{eq-dpddt}
\end{equation}

In our universe, it is reasonable to assume that $f(\tau)$ has a
broad (at least Gyr-scale) peak at some $\tau_{\rm peak}\sim O(10$
Gyr$)$, since at early times, there was no structure, and at late
times, there will be no free energy.  In any case, there will be no
features in $f(\tau)$ that can possibly compete with the
exponential factor in Eq.~(\ref{eq-dpddt}), which suppresses the
probability of late-time observations at a characteristic rate set by
the microphysical scale $H_{\rm big}$.

For example, with Planckian $H_{\rm big}\sim O(1)$, there are at any
time $t$
\begin{equation}
  \frac{f(13\, {\rm Gyr})}{f(13.7\, {\rm Gyr})} 
  \exp(10^{60})\approx \exp(10^{60})
\end{equation}
observers who live 13 Gyr after their local big bang, for every
observer like us.  Thus, the probability of seeing a 13.7 Gyr old
universe with a 2.7 K background temperature is vanishingly small
compared to the observation of a warmer CMB and a somewhat younger
universe.  

This obviously contradicts experiment.  Note that the probability for
what we do see is so small that our observations so far are, by any
standards applied in science, perfectly sufficient to rule out the
theory---or in this case, the measure.

\subsection{Explicit conflict with observation}
\label{sec-mod}

A possible objection to the above analysis is the fact that $\tau$,
the time since the big bang in our bubble, is not a physical
observable.  Therefore, following Tegmark~\cite{Teg05}, let us verify
explicitly that the youngness paradox manifests itself in the
probability distribution for physical observables.

Some observational consequences of the youngness pressure in this
measure were described more than ten years ago by Linde, Linde, and
Mezhlumian \cite{LinLin96}. There, the authors note that the proper
time measure predicts that we are living at the center of an
underdense region they refer to as an ``infloid''. The effect they
discuss arises because regions which spend less time in slow roll
inflation (hence regions which reheat sooner and therefore are
underdense) are rewarded.  We focus here on a different effect which
is more clearly in conflict with observation: the fact that typical
observers see a different temperature than we do.

The probability distribution for the temperature is
\begin{equation}
{dp \over dT} = \int d\tau {dp \over d \tau} g(T |\tau)~,
\end{equation}
where $g(T |\tau)$ is the probability distribution for temperatures at
a fixed FRW time.  For temperatures not too far from the average
value,
\begin{equation}
  g(T |\tau) \propto {1 \over T_{\rm av}(\tau)} \exp 
  \left[
    - 10^{10} \left({T - T_{\rm av}(\tau) \over T_{\rm av}(\tau)} \right)^2
  \right]~,
\end{equation}
where $T_{\rm av}(\tau)$ is the average temperature at time $\tau$,
and the factor of $10^{10}$ arises due to the magnitude of the density
perturbations.  The probability distribution becomes
\begin{equation} {dp \over dT} \propto \int d\tau 
{f(\tau)  \over T_{\rm av}(\tau)}
\exp\left[-3
    H_{\rm big} \tau - 10^{10} \left({T - T_{\rm av}(\tau) \over
        T_{\rm av}(\tau)} \right)^2 \right]~.
\end{equation}

For the moment, let us ignore observations occurring before 10 Gyr,
because for early enough times these formulas will break down. For the
times under consideration, the average temperature satisfies
\begin{equation} {T_{\rm av}(\tau_1) \over T_{\rm av}(\tau_2)} =
  \left({\tau_2 \over \tau_1}\right)^{2/3}~.
\end{equation}
The probability distribution for temperature becomes
\begin{widetext}
\begin{equation}
{dp \over dT}  \propto 
\int_{10 ~{\rm Gyr}} d\tau\, \tau^{2/3} f(\tau)  
\exp \left[ - 3 H_{\rm big} \tau 
- 10^{10}\left( \left(T \over 3.3 K \right) \left(\tau \over 10
~{\rm Gyr}\right)^{2/3} - 1 \right)^2 \right]~.
\end{equation}
\end{widetext}
The dominant factor in the integrand is $\exp(- 3 H_{\rm big} \tau )$,
since this factor varies over the microphysical time scale $H_{\rm
  big}$, while the other factors vary on much larger time scales.
Thus the integral is dominated by the lower limit.  So the probability
distribution for the temperature, once fluctuations are taken into
account, is just equal to the distribution at the early time cutoff.
Dropping a $T$-independent normalization factor, we find
\begin{equation}
{dp \over d T} \propto g(T |\tau = 10 ~{\rm Gyr}) \propto \exp \left[
 - 10^{10} \left({T - 3.3 K \over 3.3 K} \right)^2
   \right]~.
\label{eq-uf}
\end{equation}
It is easy to see that this prediction is ruled out at great
confidence by our observation that $T=2.7$ K.  The conflict only
becomes worse as the early time cutoff is reduced.

\subsection{Exact treatment of the bubble geometry}
\label{sec-bubble}
In this subsection, we will improve on the above analysis by taking
into account the actual dynamics and shape of bubble walls.  Our
treatment will clarify the extent to which the square-bubble
approximation is justified, and confirm that the youngness paradox
arises in the proper-time measure.

Inside a single bubble, as before we define a function $f(\tau)$ giving
the number of observations per unit comoving volume per proper time,
\begin{equation}
  dN^{(1)} = f(\tau)\, d\tau\, d V_c 
  = f(\tau)\, 4 \pi \sinh^2\xi ~d\xi\, d\tau ~.
\end{equation}
To get the total number of observations, $dN$, at given $(\tau, \xi)$,
we must sum over all bubbles.  We can organize this sum in terms of
the time $t_{\rm nuc}<t$ when each bubble was nucleated. Note that
given the coordinates $(\tau, \xi)$ inside the bubble, there is an
upper limit $ t_{\rm nuc}^{\rm max}(\tau, \xi)$ on the nucleation time
so that the region of interest can be produced before the time
$t$. This relationship was derived in Sec.~(\ref{sec-exact}). The sum
becomes
\begin{equation}
  {dN \over d\tau d\xi}=\int_{0}^{ t_{\rm nuc}^{\rm max}} d t_{\rm nuc}
  \left.\frac{dn}{dt}\right|_{t_{\rm nuc}} {dN^{(1)} \over d\tau d\xi}~,
\label{eq-gg}
\end{equation}
where the bubble production rate $dn/dt$ is still given by
Eq.~(\ref{eq-nour}).  Plugging in, we get
\begin{equation} {dN \over d\tau d\xi} = \int_{0}^{ t_{\rm nuc}^{\rm
      max}} d t_{\rm nuc}\, C \exp \left(3 H_{\rm big} t_{\rm nuc}
  \right) f(\tau)\, 4 \pi \sinh^2\xi
\end{equation}
where, as derived in (\ref{eq-app}),
\begin{equation} 
  t_{\rm nuc}^{\rm max} = t - \tau-H_{\rm
    big}^{-1}\left( \xi + e^{- \xi} - 1 + ... \right)~.
\end{equation}
Performing the integral and dropping constant factors, we get
\begin{eqnarray}
  & &{dN \over d\tau d\xi} = \left[
    e^{3 H_{\rm big} t^{\rm max}_{\rm nuc}} -1 \right]
  ~ f(\tau) \sinh^2 \xi \\ \nonumber 
  &=&\left[e^{3 H_{\rm big} 
      \left[t - \tau -H_{\rm big}^{-1}(\xi +e^{-\xi} - 1 + ...)\right] 
   }- 1 \right]
  f(\tau)\, 
  \sinh^2\xi 
\end{eqnarray}
Taking the limit $t \to \infty$, we can ignore the ``$-1$'' coming
from the lower limit of integration; in this limit the $t$ dependence
is only an overall multiplicative factor which vanishes upon
normalization. Thus we obtain a simple formula for the probability
distribution
\begin{equation}
{d N \over d \xi d \tau} = f(\tau)  e^{-3 H_{\rm big} \tau}~ 
  \sinh^2\xi e^{-3 (\xi+e^{-\xi} - 1 + ...)}
\end{equation}
The striking feature of this probability distribution is that it
factorizes into a function of the spatial coordinate $\xi$ times a
function of the FRW time $\tau$. This is exactly true, because the
``$\ldots$'' appearing in the formula is a function of $\xi$ only.

The distribution as a function of $\tau$ is given by
\begin{equation}
\frac{dp}{d\tau}\propto f(\tau) e^{-3 H_{\rm big} \tau}~.
\label{eq-eyp}
\end{equation}
This distribution is exactly the same as Eq.~(\ref{eq-dpddt}), which
was derived in the square bubble approximation.  So the youngness
paradox appears in exactly the same way in the true geometry.

The spatial distribution of observers at a fixed FRW time $\tau$ is
\begin{equation}
\frac{dp}{d\xi}\propto \sinh^2\xi \exp\left[-3 (\xi+e^{-\xi} - 1 +
  ...)\right]
\end{equation}
This distribution peaks at $\xi$ of order one, and falls exponentially
for large $\xi$. So most observers live within a few curvature radii
of the ``center of the universe.''  The center is defined by the
geodesic piercing the bubble nucleation point.

\subsection{Why did the square bubble approximation work?}
\label{sec-location}
 The main effect of the correctly computed probability distribution 
over $\xi$ is to allow only an effective comoving volume
\begin{equation}
  V_{\xi, {\rm eff}}=4\pi\int_0^\infty d\xi
  \sinh^2\xi \exp\left[-3(\xi-1+e^{-\xi})\right]\approx 15.75
\end{equation}
to contribute for every bubble.

The above result could not have been computed in the square bubble 
approximation, but it explains why that approximation worked for 
computing the {\em temporal\/} distribution of observers.  The point 
is that in a large regime, it is possible to identify the {\em finite\/} 
spatial region containing typical observers with the finite flat patch 
of ``new vacuum'' inserted by hand in the square bubble approximation.  

This identification is not a true match, because of the different
spatial curvature. But during and after inflation, there is a long
period where curvature is negligible and the scale factor would be the
same function of time in a flat FRW universe.  If the observations
contributing to the measure occur in this regime, then the use of
spatially flat time-slices in the square-bubble approximation will be
legitimate.

The effective physical volume at large FRW time $\tau$ is $15.75\,
a^3(\tau)$.  During inflation, $a(\tau)=H_{\rm inf}^{-1}\sinh H_{\rm
  inf}\tau$.  To match this to the square bubble physical volume, $V_y
H_{\rm inf}^{-3} \exp\left(3H_{\rm inf} \tau \right)$, at $\tau \gg
H_{\rm inf}^{-1}$, requires the choice
\begin{equation}
  V_y=V_{\xi, {\rm eff}}/8\approx 2~.  
\label{eq-vy}
\end{equation}

Note that in a problem involving different types of bubbles, the
physical volume of a new bubble will be of order $H_{\rm inf}^{-3}$.
Generically $H_{\rm inf}$ will be smaller than the outside Hubble
constant.  If we took Eq.~(\ref{eq-vy}) literally, the square bubble
approximation would involve replacing a large number of outside Hubble
volumes with the new vacuum.  This contradicts the geometric fact that
asymptotically, the new bubble takes up the comoving volume occupied
by only one outside Hubble volume at the time of nucleation.  Of
course, the choice of $V_y$ dropped out of ratios, so it could be
reduced without affecting relative probabilities.

In any case, while the square bubble approximation turned out to be a
useful shortcut under the above assumptions, it is just as simple, and
much more reliable, to use the exact geometry, as encoded in
Eq.~(\ref{eq-exact}), to compute probabilities.

\section{Modifications of the proper time measure}
\label{sec-fixes}

Obviously, the result that practically all observers live at a much
earlier time, and see a very different universe, than we do, is fatal
for the proper time measure.  Perhaps the measure can be modified in
some way, so as to avoid this problem? 

\subsection{Don't ask, don't tell}

Linde advocates a simple resolution to the youngness paradox in
Ref.~\cite{Lin06} (see also references therein).  One should simply
not ask how long after reheating the typical observers form, but
merely compute the rate at which reheating hypersurfaces of different
inflating vacua are produced.  This restriction has a number of
problems.  If we cannot ask about the temperature measured by a
typical observer, the measure is not complete.  Moreover, if we cannot
ask about observers, then we cannot count them, and so we cannot
condition on their number.  This would eliminate the anthropic
solution to the cosmological constant problem.  And finally, as noted
in Ref.~\cite{Lin07}, this restriction does not fully solve the
youngness problem in any case.  It merely confines the problem to
effects before reheating.  In particular, it gives overwhelming weight
to vacua with a shorter period of inflation, and thus predicts a wide
open universe.  Thus, a different modification is needed.

A general idea for such a fix was outlined in Ref.~\cite{Lin07}: ``One
should compare apples to apples, instead of comparing apples to the
trunks of the trees.''  In other words, we should assign a correction
factor $e^{3 H_{\rm big} \Delta t_i}$ to the probability $p_i$ for the
observation $O_i$, where $\Delta t_i$ is the amount of time it takes
to produce such an observation, in some relative sense to be defined
below.  The corrected relative probabilities are thus:
\begin{eqnarray}
  \frac{P(O_1)}{P(O_2)} &=& 
  \frac{p(O_1)}{p(O_2)} \exp[3 H_{\rm big} (\Delta t_1-\Delta t_2)]
  \\ \nonumber
  &=&\lim_{t\to\infty}\frac{N_1(t)}{N_2(t)} 
  \exp[3 H_{\rm big} (\Delta t_1-\Delta t_2)]~.
\label{eq-fix}
\end{eqnarray}
Compared to Eq.~(\ref{eq-prob}), this bolsters mature folk like
ourselves, by compensating for the enormous volume growth that
Boltzmann babies can take advantage of.

No general, sharp definition of $\Delta t_i$ was attempted in
Ref.~\cite{Lin07}, where explicit calculations were carried out only
for a model containing two vacua with different lengths of inflation;
$\Delta t_i$ was defined to be the duration of inflation in each
vacuum.  While Ref.~\cite{Lin07} claimed that the procedure also
resolves other aspects of the youngness paradox, such as the
overwhelming probability for a hotter universe, it offered no
definition of $\Delta t$ in that context, nor did it display an
explicit computation of the corrected probability.

In fact, we have been unable to come up with a general definition of
$\Delta t$ that succeeds in fixing the youngness problem.  This does
not mean that it cannot be done.  But perhaps it will help sharpen the
challenge if we discuss a few proposals that may come to
mind.\footnote{We thank Andrei Linde for discussions that influenced
  some of the definitions explored below.  However, we make no claim
  that any of them reflect his views accurately.}

\subsection{Spatial averaging}

The prediction that we should observe a warmer CMB temperature arose
from the fact that it takes longer to produce observers who see low
CMB temperatures.  To get a more reasonable probability distribution
for the CMB temperature, we want to eliminate the enormous cost of
waiting for the universe to cool off.  It seems reasonable to assign
$\Delta t(T)$ as the amount of FRW time until the average background
temperature is $T$. Since we will only be comparing observations
within the same bubble and after reheating, an additive constant in
$\Delta t$ is unimportant and so we can start our clock at any time we
like.

We explore this proposal mostly because it seems like the most
straightforward and naive fix.  In fact, it is unclear how the above
definition would generalize to observables that can take on the same
value at very different times.  Even for the temperature, small
perturbations render the relation between its average value and a
particular time slice ambiguous at the $10^{-5}$ level---much too
large to define $\Delta t$ with the required Planckian precision.  We
will disregard all these issues, since the modification fails even in
the idealized special case we consider.

For temperatures and times close to those we observe, the average
temperature on $\tau=$ const slices satisfies
\begin{equation} {T(\tau_1) \over T(\tau_2)} = \left({\tau_2 \over
      \tau_1}\right)^{2/3}~.
\end{equation}
Thus, $\Delta t(T)$ is given by
\begin{equation}
  \Delta t(T) = (13.7~ {\rm Gyr}) 
  \left( 2.7~{\rm K} \over T \right)^{3/2}~.
\label{eq-classical}
\end{equation}

The modification fails because it is comparatively easy to find
deviations from the average temperature.  To see this, let us begin by
considering a further idealization: Let us exclude fluctuations of
$T$.  In other words, we will assume that the CMB temperature, at
fixed $\tau$, is given everywhere precisely by the same value.  

With this additional idealization, the modification actually works!
There is now a one-to-one correspondence between $\tau$ and $T$, so we
can use Eq.~(\ref{eq-dpddt}) to obtain the (unmodified) probability
distribution for $T$:
\begin{equation} {dp \over dT} = {dp \over d\tau} {d\tau \over dT}
  \propto f(\tau(T)) {d\tau \over dT} \exp \left[-3 H_{\rm big}
    \tau(T) \right]
\label{eq-cluf}
\end{equation}
where $f(\tau)$, as before, is the rate of observations per comoving
volume per unit time per bubble. The quantity $f(\tau(T)) {d\tau \over
  dT} $ is naturally identified as $f(T)$, the rate of observations
per comoving volume per unit background temperature per bubble.  Using
the previous formula relating time to temperature, we find
\begin{equation} {dp \over dT} \propto f(T) \exp\left[- 3 H_{\rm big}
    \cdot (13.7~{\rm Gyr}) \left(2.7~{\rm K} \over T\right)^{3/2}
  \right] ~.
\end{equation}

Still working in the idealization of exactly homogeneous background
temperature, let us now compute the modified probability distribution
for temperature. It is
\begin{equation} {dP \over dT} \propto f(T) \exp \left[-3 H_{\rm big}
    (\tau(T) - \Delta t(T)) \right]
\end{equation}
We have defined $\Delta t$ so that the exponent is zero, so the modified 
probability distribution for temperature is simply proportional to the 
number of observations at each temperature,
\begin{equation}
{dP \over dT}  \propto f(T)~.
\label{eq-dream}
\end{equation}
This answer seems intuitive and has no youngness problem. (See,
however, the discussion at the end of Sec.~\ref{sec-anticipate}.)

Once we allow for fluctuations of the temperature, $\Delta t(T)$ can
still be defined in terms of the average temperature.  But our recipe
for repairing the probabilities no longer works.

Now the starting point is the (unmodified) probability distribution
obtained in Sec.~\ref{sec-mod}, Eq.~(\ref{eq-uf}).  After applying
Eq.~(\ref{eq-fix}), with $\Delta t(T)$ given by
Eq.~(\ref{eq-classical}), we obtain the modified distribution
\begin{equation} {dP \over d T} = \exp \left[3 H_{\rm big}\Delta t(T)
  \right] g(T |\tau = 10 ~{\rm Gyr})
\end{equation}
Using 
\begin{equation}
  \Delta t(T) = (10 ~{\rm Gyr}) \left( 3.3 K \over T \right)^{3/2}~,
\end{equation}
and assuming Planckian $H_{\rm big}$, we get
\begin{equation}
{dP \over d T} = \exp \left[ 10^{61}  \left( 3.3 K \over T
  \right)^{3/2}
 - 10^{10} \left({T - 3.3 K \over 3.3 K} \right)^2
   \right]
\end{equation}

The temperature is now driven to the {\em lowest\/} possible value.
It is still favorable to live early, and because of the primordial
density fluctuations, it is not all that hard to find an anomalously
cool region even at early times.  Our modification factor rewards us
for this as if we had honestly waited until the average temperature
becomes so low.  Thus, it overcompensates.

This new distribution is also ruled out, at enormous confidence level,
by our observation of 2.7 K.

\subsection{Waiting for the first time}

Another possibility is to define $\Delta t_i$ for the observation of
type $O_i$ as the time it takes the universe, starting from the
beginning of time (the slice $\Sigma_0$), to produce the first such
observation.

Thus defined, $\Delta t_i$---and hence, the corrected
probabilities---will depend on the initial conditions on $\Sigma_0$.
This dependence may be mild, and in any case we can see no reason why
probabilities (at least for some observables) should not depend on the
initial conditions of the universe.  However, if we define $\Delta
t_i$ as the time when $N_i$ jumps from 0 to 1, then it will also
depend on accidents of the semiclassical evolution at early times,
such as the time when a particular tunneling event happens to take
place, and we would not be able to compute it directly from the
theory.

This problem can be resolved by defining $\Delta t_i$ to be the time
when the expectation value $\langle N_i(\Delta t)\rangle$ becomes
1.\footnote{More generally, one could consider defining $\Delta t$ to
  be the time when the expectation value reaches some fixed value
  $N_{\rm min}$.  In Eq.~(\ref{eq-vp}) below, the small volume limit
  is equivalent to taking $N_{\rm min}\to\infty$ at fixed $V$.}  This
still depends on initial conditions but can be computed from the
theory in the semiclassical regime.

However, this definition conflicts with an important property of
probabilities.  Consider the special case that $O_1$ and $O_2$ are
mutually exclusive outcomes of an experiment.  For example, outcome
$O_1$ ($O_2$) may be up (down) when the spin of a single electron is
measured by a man in a penguin suit.  In general there may be
additional possible outcomes $O_3,\ldots$, but in any case, it must be
true that
\begin{equation}
p_1+p_2=p_{12}~,
\end{equation}
where $p_{12}$ is the probability for the outcome ``1 or 2''.  Indeed,
this property will be satisfied by the original probabilities defined
in Eq.~(\ref{eq-prob}).

However, $\Delta t_{12}<\Delta t_i$, $i=1,2$, because the expected
time when ``1 or 2'' is first observed is simply the time when the
experiment is first likely to be performed.  This is sooner than the
expected time when, say, 1 is first observed, since the very first
experiment can only have one outcome.  Therefore, the corrected
probabilities do not add up correctly:
\begin{eqnarray}
  P_1+P_2&=&
  p_1 e^{3 H_{\rm big} \Delta t_1} +p_2 e^{3 H_{\rm big} \Delta t_2}
  \\ \nonumber
  &>&(p_1+p_2) e^{3 H_{\rm big} \Delta  t_{12}} = P_{12}~,
\end{eqnarray}
Another way of saying this is that we can change the total probability
for a set of alternative outcomes by whether we view the alternatives
separately and add probabilities, or group the alternatives together
and directly compute the probability for this compound outcome.  This
is clearly absurd.

A particularly simple and striking result obtains if we assume initial
conditions that are already in the stationary regime.  Then
\begin{equation}
  \langle N_i(t)\rangle  = V p_i (e^{3 H_{\rm big} t}-1)~.
\label{eq-nstat}
\end{equation}
The uncorrected probabilities $p_i$ are dynamically determined by the
attractor behavior; the overall scaling $V$ depends on the volume of
$\Sigma_0$.  We find for the correction factor
\begin{equation}
e^{3 H_{\rm big} \Delta t_i}= \frac{Vp_i+1}{Vp_i}~,
\end{equation}
and hence
\begin{equation}
\frac{P_i}{P_j}=\frac{V p_i + 1}{V p_j+1}.
\label{eq-vp}
\end{equation}
In the large volume limit, there is no correction and the youngness
paradox persists.  For finite $V$, the corrected probabilities do not
obey $P_1+P_2=P_{12}$.  In the limit $V\to 0$, all alternatives become
equally likely, $P_i/P_j=1$, no matter how they were defined!

\subsection{Growing together}

A different definition for $\Delta t_i$ may be motivated by another
quote from Ref.~\cite{Lin07}: $\Delta t_i$ is ``the time when the
stationary regime becomes established'' for the observation $O_i$.
Mathematically, we may attempt to capture this idea as follows.  At
late times, we know that the number of observations of any type will
grow as $e^{3 H_{\rm big} t}$, so $\dot N_i/N_i\to 3 H_{\rm big} $.
At any finite time, there will be a small correction to this time
dependence, so we may define $\delta t_i$ as the earliest time when
\begin{equation}
  \left|1-\frac{\dot N_i(t_i)}{3 H_{\rm big} N_i(t_i)}\right|
  \leq \epsilon~.
\end{equation}

It would seem arbitrary to specify an particular small deviation
$\epsilon$ beyond which we consider the stationary regime established.
Therefore, let us take the limit $\epsilon\to 0$.  In this limit each
$\delta t_i$ contains the same additive divergence, $-\log \epsilon$,
which we discard and define
\begin{equation}
\Delta t_i=\lim_{\epsilon \to 0}\delta t_i + \log\epsilon~.
\end{equation}

To see that this measure does not work, let us focus again on the CMB
temperature in our own vacuum.  For the sake of argument, suppose that
no observers exist prior to some cutoff FRW time, say, $\tau_{\rm
  min}=10$ Gyr.  By the results of Sec.~\ref{sec-mod}, for any finite
geodesic time $t$, practically all observations of {\em any\/} value
of $T$ are made within a time of order $H_{\rm big}^{-1}$ after
$\tau_{\rm min}$.  Therefore, to accuracy $H_{\rm big}^{-1}$,
$|1-{\dot N_T}/(3 H_{\rm big} N_T)|$ will drop below $\epsilon$ at the
same time $t$, for any temperature $T$.  Hence, $\Delta t_T$ is
independent of $T$ to this accuracy.\footnote{A finite width $dT$ is
  implicit; see Eq.~(\ref{eq-density}) and the footnote thereafter.
  We use $N_T$ as a short notation for $\frac{dN}{dT} dT$. Our
  conclusion becomes strictly true only in the $\epsilon \to 0$ limit,
  when the total volume of the FRW cutoff surfaces between $\Sigma_0$
  and $\Sigma_t$ becomes large enough to contain all possible values
  of $T$.}  Therefore, our ``modification'' does not in fact change
relative probabilities at all.

To complete this argument, we should now take the cutoff FRW time,
$\tau_{\rm min}$, earlier and earlier, until it is removed altogether.
However, this introduces only information about early universe physics
into our modification of the proper time measure.  It cannot possibly
restore a reasonable probability distribution for the CMB temperature
measured by observers in the present era.

It is interesting that like ``spatial averaging'', the present
modification {\em would\/} have worked for (fictitious) observables
that are in one-to-one correspondence with the FRW time.  In this
case, it would not have been true that at any time $\tau$, there is a
nonzero amplitude for any temperature $T$.  Instead,
\begin{equation}
  N_{T_1}(t)=
  \frac{f(T_1)}{f(T_2)}N_{T_2}(t+\tau(T_2)-\tau(T_1))~,
\end{equation} 
and hence,
\begin{equation}
  \frac{\dot N_{T_1}}{N_{T_1}} (t)= 
  \frac{\dot N_{T_2}}{N_{T_2}}(t+\tau(T_2)-\tau(T_1))~.
\end{equation}
Then we would have found
\begin{equation}
\Delta t_1-\Delta t_2=\Delta t(T_1)-\Delta t(T_2)~,
\end{equation}
and we would have recovered the intuitive result of
Eq.~(\ref{eq-dream}).

Apparently, the problem with both of these modifications is that the
{\em value\/} of an observable does not give us enough information
about the FRW {\em time\/} when the observation is made---but this is
precisely the time we would like to use for $\Delta t$.  This
motivates our final attempt at modifying the proper time measure, in
which $\Delta t$ is defined not as a function of observables, but
directly as a function of the time when the observation is made,
regardless of its outcome.

\subsection{Anticipation}
\label{sec-anticipate}

Instead of tying $\Delta t$ to a specific observable, we can go back
and fix the time shift directly for the geodesics.  Effectively, it is
like ``anticipating'' all observations that will happen in a given
bubble.  More precisely, let us project every observation $O_i$ inside
a bubble back to the most recent bubble wall along the geodesics of
the congruence, and count it toward $N_i(t)$ as soon as the relevant
portion of the wall lies below $\Sigma_t$.  This amounts to choosing
$\Delta t_i$ to be the geodesic time between the domain wall and the
observation.  It is not difficult to see that this choice eliminates
any pressure to make observations very early, and that it reduces to
the $\Delta t$'s used in the specific example of
Ref.~\cite{Lin07}. However, in general it suffers from two major
problems.

First, it is not sharply enough defined.  The prescription involves
projecting onto domain walls.  These objects have an inherent
thickness, which can be microscopic, but need not be Planckian. This
is a problem because we need a proposal which is well-defined at the
length scale $H_{\rm big}^{-1}$, which may be Planckian.  Moreover, an
approximately defined object like a domain wall has no place in a
fundamental definition of probabilities for all observations.  There
is a smooth interpolation between objects that appear obviously
recognizable as domain walls, and general field configurations.  (This
objection could be raised also against other measures that involve
domain walls in their definition, such as Ref.~\cite{GarSch05}.)

On the observational side, the projection method suffers from the
``Boltzmann brain'' problem~\cite{Pag06,BouFre06b}. The reason is that we
are now completely indifferent to when observations inside the bubble
are made. By the results of Sec.~\ref{sec-location}, we may focus on a
single comoving volume at the center of any metastable de~Sitter
bubble with sufficiently small cosmological constant (such as,
presumably, our own vacuum).  An infinite number of observers are
formed at late times in this volume, due to rare thermal
fluctuations~\cite{Pag06}.  All of these Boltzmann brains will be
projected back, and so will dominate over other observers.  Thus, with
probability 1, we should be Boltzmann brains, which is in conflict
with observation~\cite{DysKle02}.  (Alternatively, we can interpret
this infinity as telling us that projection defeats the most basic
purpose of the measure, which is to regulate the infinities occurring
in eternal inflation.)

Mathematically, the Boltzmann brain problem shows up as follows.  The
effect of the ``anticipation'' modification is to render the
temperature distribution apparently well-behaved: we have finally
succeeded in producing the hoped-for Eq.~(\ref{eq-dream}).  But this
equation is a poisoned chalice: it is not as harmless at it looks.
Boltzmann brains arise at a fixed rate per unit time and unit physical
volume, not per comoving volume.  Thus, the comoving observer density
$f(\tau)$ grows exponentially with the scale factor at extremely late
times, so $f(T)$ diverges at the Hawking temperature of the de~Sitter
space.

\acknowledgments We are grateful to Andrei Linde for extensive
discussions.  This work was supported by the Berkeley Center for
Theoretical Physics, by a CAREER grant of the National Science
Foundation, and by DOE grant DE-AC03-76SF00098.

\bibliographystyle{board}
\bibliography{all}

\end{document}